\documentclass[twocolumn,prb,amsmath,amssymb,showpacs,superscriptaddress,floatfix]{revtex4}

\usepackage{graphicx}

\begin{document}

\title{Appearance of Schr\"odinger Cat States in the Measurement Process}

\author{G.B.\ Lesovik}
\affiliation{L.D.\ Landau Institute for Theoretical Physics RAS,
117940 Moscow, Russia}

\author{A.\ Lebedev}
\affiliation{L.D.\ Landau Institute for Theoretical Physics RAS,
117940 Moscow, Russia}

\author{G.\ Blatter}
\affiliation{Theoretische Physik, ETH-H\"onggerberg, CH-8093
Z\"urich, Switzerland}

\date{\today}

\begin{abstract}
Although quantum mechanics is a mature theory, fundamental
problems discussed during its time of foundation have remained
with us to this day. These problems are centered on the
problematic relation between the quantum and classical worlds. The
most famous element is the measurement problem, i.e., the
measurement of a quantum system by a classical apparatus, and the
concomitant phenomena of wave packet reduction, the appearance of
probability, and the problems related to Schr\"odinger cat states.
A fundamental question in this context is whether quantum
mechanics can bootstrap itself to the classical world: is quantum
mechanics self-consistent, such that the measurement process can
be understood within quantum mechanics itself, or does this
process require additional elements from the realm outside of
traditional quantum mechanics? Here, we point to a problematic
aspect in the traditional Schr\"odinger cat argument which can be
overcome through its extension with a proper macroscopic
preparation device; the deliberate creation of a cat state and its
identification then turns into a non-trivial problem requiring the
determination of the evolution of a quantum system entangled with
a macroscopic reservoir. We describe a new type of wave-function
correlator testing for the appearance of Schr\"odinger cat states
and discuss its implications for theories deriving the wave
function collapse from a unitary evolution.
\end{abstract}


\maketitle

\section{Introduction}
The most widespread interpretation of quantum mechanics follows a
pragmatic approach, starting with the request not to attribute a
deep meaning to the wave function but immediately go over to
probabilities \cite{born,pauli}. The randomness exists as a
postulate, the Born rule defines the probability $\rho = |\Psi|^2$
from the wave function $\Psi$, and the von Neumann projection
postulate for the wave function or density matrix tells us how to
deal with repeated measurements and calculate correlators.
Schr\"odinger cats do not exist (as no meaning is attributed to
the wave function except for providing a tool to determine
probabilities) and the separation between the quantum and the
classical worlds appears to be arbitrary.

Within this pragmatic approach, the collapse of the wave function
in the measurement process appears as a particularly mysterious
step: Consider an initial state
\begin{equation}
   |\Psi\rangle_{t=0}=|\varphi\rangle \otimes |{\bf M}\rangle
   \label{init}
\end{equation}
describing the microscopic quantum system ($|\varphi\rangle$) and
the macroscopic measurement device ($|{\bf M}\rangle$, a reservoir
in some metastable state). Following standard arguments
\cite{vonneumann}, the measurement process involves two related
operators $\hat{m}$ and $\hat{M}$ (with eigenstates $|m_n\rangle$
and $|{\bf M}_n\rangle$) acting on the microscopic quantum system
and the macroscopic meter, respectively. After the measurement
process, the systems' state can be expressed in the basis
$|m_n\rangle$ and $|{\bf M}_n\rangle$ of the measurement
operators, where the eigenvalue $M_n$ describes the meter reading
of the apparatus,
\begin{equation}
   |\Psi\rangle_{t=t_{\rm measure}}
   =\sum_n \varphi_n(t) |m_n\rangle \otimes |{\bf M}_n\rangle,
   \label{cat}
\end{equation}
and with the coefficients $\varphi_n = \langle m_n|\varphi\rangle$
related to the initial state $|\varphi\rangle$ of the quantum
system. It is the superposition of orthogonal macroscopic meter
states (allowing for a classical reading) which is the problematic
aspect of the cat state (\ref{cat}). Its {\it ad hoc} resolution
then invokes a collapse of the wave function by requiring a unique
classical meter reading \cite{landaulifschitz}, resulting (with
probability $|\varphi_{n^*}|^2$) in the collapsed wave function
\begin{equation}
   |\Psi\rangle_{t={t_{\rm measure}}}
   =|m_{n^*}\rangle \otimes |{\bf M}_{n^*}\rangle
   \label{coll}
\end{equation}
and the systems' localization in the state $|m_{n^*}\rangle$ after
the measurement.  Hence, we conclude that the wave function
collapse is not realized within the frame given by ordinary
quantum mechanics and the question arises why this is so, is this
only a technical problem due to the macroscopic nature and the
complexity of the measurement apparatus or is some hidden new
physics required for its resolution?

Here, we wish to pursue the idea that the wave function collapse
could be explained within the framework of quantum mechanics
alone, i.e., based on a unitary evolution of the initial state
(\ref{init}). We are immediately confronted with two problems:
{\it i)} combining linearity with unitary evolution seemingly
produces Schr\"odinger cat states (\ref{cat}) rather than
localized states of the type (\ref{coll}); and {\it ii)} the
unitary evolution to a localized state (\ref{coll}) deprives us
from the element of randomness in the theory and the question
poses itself how the concept of probability (with the correct Born
rule) can be recovered. Below, we will argue that the {\it
traditional} derivation of the Schr\"odinger cat state suffers
from the problem of inappropriate initial conditions. We will show
that the appearance of a Schr\"odinger cat state requires the
presence of a superposed macroscopic preparation device already to
begin with; the deliberate creation of a Schr\"odinger cat state
then becomes a non-trivial issue. In fact, as of now it is unclear
whether a carefully prepared state indeed develops into a
Schr\"odinger cat state under unitary evolution: the confirmation
of the appearance of a Schr\"odinger cat state requires
calculation of the unitary time evolution of the entangled ground
state comprising the particle and a macroscopic preparation
device; in principle, this could be done using standard tools
\cite{leggett_rev}, but no answer has been given so far. A
dramatic simplification of this task is expected when the
environment states are averaged over, e.g., at finite
temperatures; unfortunately, using standard tools, the information
about the presence or absence of Schr\"odinger cat states is lost
in the averaging process. Below, we introduce a new type of
wave-function correlator allowing to decide on the appearance of
Schr\"odinger cat states even after averaging and we discuss first
results obtained with the help of this technique. Also, with
respect to point {\it ii)} raised above, we will argue that the
element of randomness appears naturally through the indeterminacy
of the initial state of the macroscopic environment which appears
in the preparation of the physically correct initial quantum state
and in its measurement.

\section{Schr\"odinger Cat States}
\begin{figure}[h]
\includegraphics[scale=0.50]{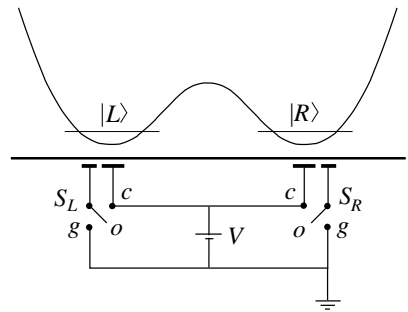}
\caption[]{Schematic (electric) setup producing a double-well
potential with suitable switches $S_{L(R)}$ for state preparation:
$c$: biased state, $o$: unbiased but polarized state, $g$:
unbiased state without polarization.} \label{fig:elec}
\end{figure}

Let us discuss first the traditional argument leading to the
appearance of Schr\"odinger cat states. To fix ideas we consider a
double-well potential with semiclassical states $|L\rangle$ and
$|R\rangle$ (particle in the left/right well, cf.\ Fig.\
\ref{fig:elec}) coupled to a macroscopic reservoir $|{\bf
M}\rangle$ serving as a meter (the macroscopic cat). The usual
Schr\"odinger cat argument assumes that the individual product
states involving the quantum system and an unpolarized reservoir
$|{\bf M}_{\rm up}\rangle$ evolve into polarized states
\begin{eqnarray}
   &&|L\rangle \otimes |{\bf M}_{\rm up}\rangle \rightarrow
   |L\rangle \otimes |{\bf M}_{L}\rangle,\nonumber \\
   &&|R\rangle \otimes |{\bf M}_{\rm up}\rangle \rightarrow
   |R\rangle \otimes |{\bf M}_{R}\rangle;
   \label{sip_evol}
\end{eqnarray}
the states $|{\bf M}_{L}\rangle$ and $|{\bf M}_{R}\rangle$ denote
the macroscopic polarized states of the detector after the
measurement. Assuming linearity, it is concluded that the state
$[|L\rangle+|R\rangle]\otimes |{\bf M}_{\rm up}\rangle$ evolves
into a Schr\"odinger cat state,
\begin{equation}
   [|L\rangle+|R\rangle]\otimes |{\bf M}_{\rm up}\rangle \rightarrow
   |L\rangle \otimes |{\bf M}_{L}\rangle+
   |R\rangle \otimes |{\bf M}_{R}\rangle;
   \label{sup_evol}
\end{equation}
it is the superposition of the macroscopic reservoir (the cat)
which is the problematic aspect of the state (\ref{sup_evol}),
cf.\ also (\ref{cat}). The underlying mistake in the above
Schr\"odinger cat argument is the assumption that the whole
procedure can be carried out with a single unitary evolution. On
the contrary, in order to realize the evolution in
(\ref{sip_evol}) the well has to be externally biased (e.g.,
through a linear potential tilting the double-well) in order to
prevent the particle from leaving the left (right) well during the
time evolution associated with the measurement; hence
(\ref{sip_evol}) involves the biased unitary evolutions
$\hat{U}_L$ and $\hat{U}_R$. On the other hand, the evolution
(\ref{sup_evol}) {\it has to} proceed without any bias and hence
involves a different unitary operator $\hat{U}_0$. We conclude
that this simplistic manner to create a cat state cannot work.

\begin{figure}[h]
\includegraphics[scale=0.35]{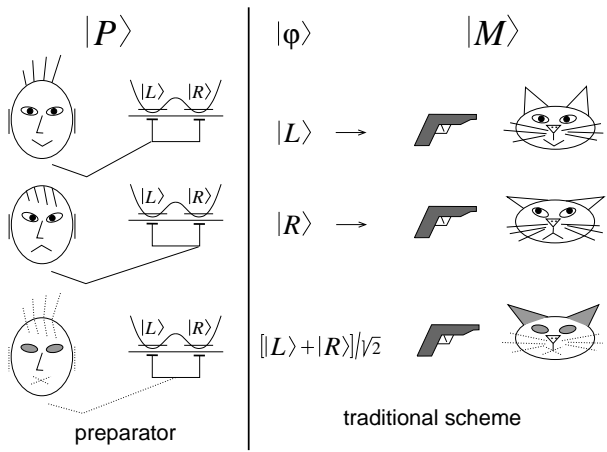}
\caption[]{Traditional (right side, involving the microscopic
  quantum system $|\varphi\rangle$ and the measurement device
  $|{\bf M}\rangle$) and new extended (with the
  macroscopic preparator $|{\bf P}\rangle$ included)
  schemes for the construction of a Schr\"odinger cat state.
  The `strange' preparator and cat in the last row are macroscopic
  superpositions of the happy and unhappy preparator/cat
  appearing in the first two rows. We argue, that in order to
  construct a cat state along the traditional argument involving
  linearity and unitarity, a superposition of the macroscopic
  preparator is required --- a final cat state requires an initial
  cat state to begin with.}
  \label{fig:cat_sk}
\end{figure}

We then consider a more sophisticated setup for the construction
of a Schr\"odinger cat state. In our second attempt we want to use
the same unbiased unitary evolution in the measurement process
(i.e., at times $t > 0$) both for the individual wave functions
($|L(R)\rangle \otimes \dots$) and for the superposed state
($[|L\rangle+|R\rangle] \otimes\dots$). We then have to be careful
how to prepare the superposed state at times $t < 0$. To be
specific, we again consider the setup shown in Fig.\
\ref{fig:elec}, with a charged particle trapped in a symmetric
double-well potential set up electrostatically. In addition, we
account for additional polarization charges $|{\bf P}_{L}\rangle$
and $|{\bf P}_{R}\rangle$ induced by a classical bias voltage $V$
--- the polarization charges plus switches define our preparation
device. We then prepare the states $|L\rangle \otimes |{\bf
P}_{L}\rangle$ and $|R\rangle \otimes |{\bf P}_{R}\rangle$ by an
appropriate biasing of the double-well potential, e.g., via
closing the switches $S_L$ and $S_R$, respectively (position $c$).
At the time $t=0$ we remove the classical bias $V$ (switches in
position $o$), hence at times $t>0$ the Hamiltonian describing the
setup is symmetric. Different from before, where a classical
external bias was entering the Hamiltonian at $t >0$, here, the
macroscopic preparation device disconnected from the classical
external bias is now a part of the system and its time evolution
has to be accounted for in the overall evolution. Next, we connect
the system to the unpolarized measuring apparatus $|M_{\rm
up}\rangle$. Assuming that large polarization charges remain on
the left (right) well, the two states will propagate under the
same unbiased unitary evolution according to
\begin{equation}
   |L\rangle\otimes|{\bf P}_{L}\rangle\otimes|{\bf M}_{\rm up}\rangle
   \rightarrow
   |L\rangle\otimes|{\bf P}_{L}\rangle\otimes|{\bf M}_{L}\rangle
   \label{mac_evol}
\end{equation}
and similar for $L\rightarrow R$. Making use of linearity, we find
that the superposed state evolves according to
\begin{eqnarray}
  &&
  [|L\rangle\otimes|{\bf P}_L\rangle
  +|R\rangle\otimes|{\bf P}_R\rangle]
  \otimes|{\bf M}_{\rm up}\rangle
  \label{mac_cat_evol}\\
  &&\rightarrow
  |L\rangle\otimes|{\bf P}_L\rangle\otimes|{\bf M}_{L}\rangle
  +|R\rangle\otimes|{\bf P}_R\rangle\otimes|{\bf M}_{R}\rangle.
  \nonumber
\end{eqnarray}
Obviously, the evolution is again towards a cat state (involving
the macroscopic meter states $|{\bf M}_{L(R)}\rangle$); however,
different from the traditional analysis, here, the original state
at $t=0$ is already a cat state as it involves the superposition
of the macroscopic preparation device $|{\bf P}_{L(R)}\rangle$,
cf.\ Fig.\ \ref{fig:cat_sk}. In the end, we could rescue the basic
construction of a Schr\"odinger cat state by moving the asymmetry
hidden in the Hamiltonian of the traditional argument into a
preparator device which now is a part of the system. The history
of this preparator then is imprinted in the initial condition of
the system. The inclusion of this preparator cannot be avoided:
removing the macroscopic polarization clouds holding the particle
in the left- or right well, e.g., via switching into the position
$g$, the evolution (\ref{mac_evol}) is no longer guaranteed (as
the particle can delocalize between the wells) and the whole
argument is spoiled. In fact, let us prepare the $t=0$ state
$|L\rangle\otimes|{\bf P}_L\rangle$ and then turn the switch $S_L$
into the position $g$ such as to let the particle evolve into a
delocalized state involving both semi-classical states $|L\rangle$
and $|R\rangle$ (with amplitudes $\nu_{L}$ and $\nu_{R}$). At the
time $t=t_{\rm p}$ our prepared state takes the superposed ($S$)
form
\begin{eqnarray}
   |\Psi_{S}\rangle &=&
   [\nu_L(t_{\rm p})|L\rangle
   \otimes|{\bf P}_{LL}(t_{\rm p})\rangle
   \label{prep} \\ && \qquad
   + \nu_R(t_{\rm p})|R\rangle
   \otimes|{\bf P}_{LR}(t_{\rm p})\rangle],
   \nonumber
\end{eqnarray}
where $|{\bf P}_{LL}(t_{\rm p})\rangle$ and $|{\bf P}_{LR}(t_{\rm
p})\rangle$ denote the polarization states of the left/right cloud
if the preparator was initially left-polarized at $t=0$. Hence,
the particle is entangled with the reservoir that is still
carrying the information of the initial state $|{\bf P}_L\rangle$;
in the standard Schr\"odinger cat argument, this state is
approximated in terms of the product state $[\nu_L |L\rangle
+\nu_R|R\rangle] \otimes |{\bf P}_{\rm up}\rangle$.

Let us measure the state of the system at the time $t_{\rm m}\geq
t_{\rm p}$ (without loss of generality, $t_{\rm m} = t_{\rm p}$):
given the initial state $|\Psi_{\rm p} \rangle \otimes |{\bf
M}_{\rm up}\rangle$, we cannot {\it a priori} predict the outcome
of this measurement knowing only the evolution
(\ref{mac_cat_evol}), as the partly polarized reservoir states
$|{\bf P}_{LL}\rangle$ and $|{\bf P}_{LR}\rangle$ {\it differ}
from the fully polarized states $|{\bf P}_L\rangle$ and $|{\bf
P}_R\rangle$. At the present moment the answer to this correctly
posed problem is unknown --- the answer may describe a wave
function localized in one well, but also, we cannot exclude an
evolution towards a $T=0$ Schr\"odinger cat state involving a
superposition of macroscopic reservoir states.

Summarizing, we have seen that the traditional method to `create'
a Schr\"odinger cat state cannot work --- on the contrary, the
initialization of such a cat state is itself a non-trivial task
involving the entanglement of the quantum system with a
macroscopic preparation device. The time evolution of the
macroscopic preparator then has to be accounted for in the further
evolution of the system. This demanding task has been solved for
some special cases \cite{leggett_rev}: Consider a particle in a
double-well potential and coupled to an ohmic environment $|{\bf
E}\rangle$ at zero temperature (we use the general term
`environment' for a reservoir which may act as a preparator, a
measurement device, or both). Starting from a polarized state
$|\Psi_L\rangle = |L\rangle\otimes|{\bf E}_L\rangle$ (i.e., a
particle prepared in the left well at $t=0$) it has been found
\cite{leggett_rev} that, depending on the coupling strength
$\alpha$, the system evolves into a collapsed state (the particle
remains in the left well) if the coupling $\alpha>1$ is strong,
while for weak coupling $0<\alpha<1$ the particle is delocalized
between the wells (quantum coherent for $0<\alpha<1/2$ and
semi-classically for $1/2<\alpha<1$). Hence, a strongly coupled
reservoir ($\alpha>1$) defines a good preparator/meter (leading to
a wave function collapse), while a weakly coupled reservoir
($0<\alpha<1$) does neither produce a measurement nor does it lead
to a cat state. The question whether and under which conditions a
carefully prepared delocalized state $|\Psi_{S}\rangle$ ($S$ for
superposition) of the type (\ref{prep}) evolves into a
Schr\"odinger cat or into a collapsed state defines a demanding
problem which has not been pursued so far; we note, however, that
such a calculation could be carried out using standard tools,
e.g., via the determination of the occupational asymmetry ${\cal
P}_\infty = [{\cal P}_{S\rightarrow L}(t) - {\cal P}_{S\rightarrow
R}(t)]_{t\rightarrow \infty}$ at asymptotically large times (here,
${\cal P}_{S\rightarrow L(R)} (t)$ denotes the probability to find
the particle in the left (right) well at time $t$).

\section{Wave Function Collapse and Density Matrix}
The task of identifying a Schr\"odinger cat state in a properly
defined setup may be simplified dramatically if the environmental
degrees of freedom could be averaged over some ensemble, e.g., a
finite temperature Gibbs ensemble. The standard route involves the
calculation of the density matrix with an averaging over the
reservoir degrees of freedom. The diagonal elements are referred
to as the probabilities to find the particle in some specific
state --- they do not tell whether the wave function has collapsed
or not. Usually, it is the vanishing of the off-diagonal matrix
elements due to decoherence which is taken as a signal for the
collapse of the wave function, a conclusion which has lead to some
dispute recently \cite{adler}. Here, analyzing a simple
counterexample \cite{lesovik_01}, we argue that the vanishing of
the off-diagonal elements in the density matrix cannot be taken as
a proof for the collapse of the particle's wave function. Consider
the wave function (we switch to coordinate representation for
convenience)
\begin{equation}
   \Psi(x,{\bf Y}) = \nu_L \delta_{x,{x_L}} \Phi_L({\bf Y}) +
   \nu_R \delta_{x,{x_R}}\Phi_R({\bf Y})
   \label{catLR}
\end{equation}
describing a particle delocalized between the left and right
states of a double-well potential, $|\nu_L|^2 + |\nu_R|^2 = 1$.
Let the reservoir states assume the properties of normalization,
$\int d{\bf Y} \Phi_{L(R)}^*({\bf Y})\Phi_{L(R)}({\bf Y}) = 1$,
and orthogonality $\int d{\bf Y} \Phi_L^*({\bf Y})\Phi_R({\bf Y})
= 0$. Calculating the off-diagonal density matrix element by
taking the average over the reservoir coordinates,
\begin{eqnarray}
   \rho_{LR} &=& \int d[{\bf Y}] \Psi^*(x_L,{\bf Y})\Psi(x_R,{\bf Y})
    \label{od-dm}\\
    &=& \nu_L^* \nu_R \int d[{\bf Y}]\Phi_L^*({\bf Y})
   \Phi_R({\bf Y}) = 0,\nonumber
\end{eqnarray}
we obtain a zero result even for coefficients $\nu_{L(R)} \neq 0$,
i.e., while the wave function is delocalized, the off-diagonal
elements of the density matrix are zero. In a more complex example
\cite{lesovik_02} where the reservoir is modelled as a fluctuating
classical force, again, the particle wave function remains
delocalized while the off-diagonal elements in the density matrix
are shown to vanish.

This inability to discuss the collapse of the wave function within
a density matrix description is related to the loss of information
in the averaging procedure over the initial states of the
reservoir. In fact, at finite temperatures, the asymmetry ${\cal
P}_\infty$ introduced above has to be averaged over the thermal
distribution of reservoir states. In contrast to the $T=0$ case,
the finite temperature version $\langle {\cal P}_\infty
\rangle_{\rm th}$ vanishes identically for all values of the
coupling $\alpha$ (we start from a $t=0$ polarized equilibrium
state). This zero result can appear in two different ways: {\it
i)} depending on the initial state $|{\bf E}_n\rangle$ of the
reservoir the probabilities ${\cal P}^{\scriptscriptstyle
(n)}_{L\rightarrow L(R)}(t \rightarrow \infty)$ to find the
particle in the left (right) well are either 0 or 1 and the final
probability $\langle {\cal P}_\infty \rangle_{\rm th}$ vanishes
only after averaging over initial states $|{\bf E}_n\rangle$; in
this case the particle becomes localized for every given state
$|{\bf E}_n\rangle$ of the reservoir. {\it ii)} For all initial
states $|{\bf E}_n\rangle$ of the reservoir, the particle remains
delocalized and ${\cal P}^{\scriptscriptstyle (n)}_{L\rightarrow
L(R)}(t \rightarrow \infty) = 1/2$; the asymmetry ${\cal
P}^{\scriptscriptstyle (n)}(t \rightarrow \infty)$ already
vanishes for each reservoir state $|{\bf E}_n\rangle$ alone.
Hence, studying the occupational asymmetry $\langle {\cal
P}_\infty \rangle_{\rm th}$ or the diagonal density matrix
elements $\rho^L_{LL} = \langle {\cal P}_{L\rightarrow
L}\rangle_{\rm th} = \sum w_n {\cal P}^{\scriptscriptstyle
(n)}_{L\rightarrow L} = 1/2$ we cannot decide whether the particle
becomes localized or not.

\section{Wave Function Correlator}
From the above discussion, we conclude that the density matrix is
insufficient to discuss a possible localization of the particle by
the reservoir at finite temperatures. This density matrix can be
considered as the most general second-order (equal time)
correlator for the wave function $\Psi$. An analytic tool
providing an answer on the issue of the collapse of the wave
function then has to involve higher-order correlators of the
$\Psi$-function. As we are going to show now, the calculation of a
fourth-order correlator allows to separate extended from localized
wave functions. Hence, in order to decide on the appearance of
Schr\"odinger cat states we propose to calculate the correlator
\begin{equation}
   \langle {\cal P}_{L\rightarrow L}(t)
   {\cal P}_{L\rightarrow R}(t)\rangle_{\rm th}
     \equiv \sum_n w_n
     {\cal P}^{\scriptscriptstyle (n)}_{L\rightarrow L}(t)
     {\cal P}^{\scriptscriptstyle (n)}_{L\rightarrow R}(t)
   \label{correlator}
\end{equation}
with $w_n$ the Gibbs weights. The vanishing of (\ref{correlator})
then tells us that the particle becomes localized; in this case,
the number of ensemble states producing delocalized cat states is
of measure zero. On the other hand, a finite value $\langle {\cal
P}_{L\rightarrow L}(t) {\cal P}_{L\rightarrow R}(t)
\rangle|_{t\rightarrow\infty} \neq 0$ points to the presence of
Schr\"odinger cat states \cite{lesovik_01}. We note that the
correlator (\ref{correlator}) used here for the identification of
cat states is reminiscent of the participation ratio
\cite{thouless} used in the problem of Anderson localization where
it serves a similar task, the separation from extended and
localized wave functions, in another context.

Using the property ${\cal P}^{\scriptscriptstyle
(n)}_{L\rightarrow L}(t) +{\cal P}^{\scriptscriptstyle
(n)}_{L\rightarrow R}(t) = 1$, we can equivalently study the
second moment $\langle {\cal P}^2_{L\rightarrow L}(t) \rangle$;
the calculation of higher moments $\langle {\cal
P}^k_{L\rightarrow L}(t) \rangle$ then allows for the
determination of the full distribution function $\Pi\,[{\cal
P}_{L\rightarrow L}]$ of the random variable ${\cal
P}_{L\rightarrow L}$, given a distribution $w_n$ of initial states
of the reservoir. Such a calculation has been carried out
\cite{lesovik_02} for the model describing a particle trapped in a
double-well potential and subject to a classical fluctuating field
$\eta(t)$ (producing a fluctuating energy difference between the
wells) with Gaussian correlations $\langle\eta(t)\eta(t') \rangle
= \Gamma \delta(t-t')$. Eq.\ (\ref{correlator}) then is replaced
by the expression
\begin{eqnarray}
   \langle {\cal P}_{L\rightarrow L}^k(t) \rangle_\eta
   \equiv \int {\cal D}[\eta(\tau)] {P}[\eta(\tau)]
   {\cal P}^k_{L\rightarrow L}([\eta],t),
   \label{correlator_phi}
\end{eqnarray}
where ${P}[\eta(\tau)]$ denotes the weight of the particular
realization $\eta(\tau)$. The result turns out quite non-trivial
with moments $\langle {\cal P}^k_{L \rightarrow L}(t\rightarrow
\infty) \rangle_\eta = 1/(1+k)$, resulting in a homogeneous
distribution of the random variable ${\cal P}_{L \rightarrow L}$.
Interpreting the two-level system $|L(R)\rangle$ in terms of a
spin variable $|\!\uparrow(\downarrow)\rangle$, we find that after
sufficiently long times an initial state $|\!\!\uparrow\,\rangle$
points into an arbitrary direction with equal probability --- for
a given realization $\eta(\tau)$ the particle is delocalized and
remains coherent, while its density matrix (after averaging over
$\eta$) has zero off-diagonal elements.

\section{Reduction from Unitary Evolution}
Let us return to the question whether the wave function collapse
could be explained within quantum mechanics itself: a scheme
fulfilling this task should describe {\it a)} the crossover from
quantum to classical behavior, e.g., as a function of size or
coupling, {\it b)} it should describe the collapse of the wave
function (reduction of the wave packet), {\it c)} it should
explain the appearance of probability, and {\it d)} it should
produce the correct probability as given by the Born rule. In our
approach it is the reservoir which plays a central role --- we
have argued that it appears unavoidably already in the preparation
stage of an experiment and it introduces the element of randomness
into the theory. On the one hand, the scheme takes the form of a
deterministic theory, with the microscopic state of the reservoir
uniquely determining the outcome of the measurement. On the other
hand, the microscopic state of the reservoir is not known in
practice (this is implied by the definition of the term
`reservoir') and this is the basic element responsible for the
appearance of randomness. We call our scheme describing the wave
function collapse (reduction) within the framework defined by the
rules of quantum mechanics (unitary evolution) and with the
reservoir providing the element of randomness as the `R from U
scheme' (reduction from unitary evolution). The scheme is
reminiscent of von Neumann's hidden variable theory, against which
he constructed two powerful arguments \cite{vonneumann}; we
briefly discuss their implication for our R from U scheme.

The first argument is about `dispersionless ensembles': an
ensemble ${\cal E}$ is called `dispersionless' if for all
hermitian operators $A$, $\langle A^2\rangle_{\cal
\scriptscriptstyle E} = \langle A \rangle_{\cal\scriptscriptstyle
E}^2$, where $\langle \dots \rangle_{\cal\scriptscriptstyle E}$
denotes the ensemble average. There are no dispersionless
ensembles \cite{vonneumann}, hence even an ensemble made from a
single state $\varphi(x)$ exhibits dispersion. Thus randomness is
an intrinsic property of quantum mechanics and there are no hidden
variables. --- In the R from U theory, the intrinsic randomness is
replaced by the indeterminacy of the initial state of the
macroscopic reservoir and hence the degrees of freedom of the
reservoir play the role of hidden variables. While in the von
Neumann argument observables are abstract operators acting within
the Hilbert space of the quantum system, in the R from U theory,
the measurement of each such observable requires the introduction
of its own measuring device and its associated Hilbert space.
Contrary to the situation studied by von Neumann where all
operators act on the same ensemble, here, the ensemble is
specified by the measurement apparatus.

The second argument deals with the Born rule: Given an initial
density matrix in the product form $\rho = \rho_\varphi \otimes
\rho_{\bf M}$ with the particle in a pure state $\rho_\varphi =
|\varphi\rangle\langle\varphi|$ and the meter in the mixture
$\rho_{\bf M} = \sum_n w_n |\Phi_n\rangle\langle\Phi_n|$, the
unitary evolution describing the measurement process produces the
result $\rho(t) = \sum w_n |\Psi_n(t)\rangle\langle\Psi_n(t)|$
with $\Psi_n(x,{\bf Y},t) = \exp(-iHt/\hbar) \varphi(x)
\Phi_n({\bf Y})$. Assuming that the reservoir state $\Phi_n({\bf
Y})$ uniquely determines the outcome of the measurement, i.e.,
$\varphi\rightarrow\phi_n$, we arrive at the result $\rho = \sum
w_n |\phi_n\otimes \Phi_n\rangle\langle\phi_n\otimes\Phi_n|$ and
the probability to measure the $n$-th meter state is given by the
probability $w_n$, independent of the initial state $\varphi(x)$
of the particle (note that here, all of the weight in $\varphi$
collapses to $\phi_n$ and not only a fraction $|a_n|^2 =
|\langle\varphi,\phi_n \rangle|^2$). --- However, we have seen
above that an initial product state $\Psi = [\sum a_n
\phi_n(x)]\Phi_n({\bf Y})$ involving a superposition of the
quantum system cannot be prepared without a preparation device,
see (\ref{prep}); the latter has to be accounted for in the
evolution of the measurement process. As argued above, the true
wave function is not a simple direct product state but takes the
form (\ref{prep}); the partially polarized reservoir states $|{\bf
P}_{LL(R)}\rangle$ have encoded into them the amplitudes
$\nu_{L(R)}$ of the initial superposed state and, in order for the
R from U scheme to work out, the outcome of the measurement should
depend on these amplitudes such as to produce the correct Born
rule.

In conclusion, we have shown that the construction of a
Schr\"odinger cat state and its identification in a measurement is
a highly non-trivial issue involving the determination of the
unitary evolution of the quantum system entangled with its
macroscopic environment. Averaging over the environmental degrees
of freedom in order to render this task manageable implies the
loss of information on the presence of cat states if standard
tools (e.g., density matrices) are used. However, going over to
higher-order wave function correlators the question on the
appearance of cat states can be decided; the possibility to
investigate the reservoir-induced collapse paves the way for the
further analysis of the consistency of the R from U scheme.

We acknowledge discussions with B.\ Braun, J.\ Fr\"ohlich, V.\
Geshkenbein, G.M.\ Graf, M.\ Sigrist, M.\ Troyer and financial
support through the Swiss National Science Foundation (CTS and
SCOPES), the NWO grant for Collaboration with Russia, the Russian
Science-Support Foundation, Forschungszentrum J\"ulich (Landau
Scholarship), the Russian Ministry of Science, and the Russian
Program for Support of Scientific Schools.

\end{document}